\documentclass[sigconf]{acmart}

\renewcommand\footnotetextcopyrightpermission[1]{} 
\usepackage{subcaption}

\usepackage[textsize=tiny,disable]{todonotes}

\usepackage{tikz}
\usetikzlibrary{arrows}
\usetikzlibrary{positioning}  
\usetikzlibrary{shadows}
\usetikzlibrary{arrows,backgrounds,calc,chains,
	matrix,positioning,shapes,shapes.geometric,
	shapes.arrows, decorations.pathmorphing, decorations.pathreplacing}
\usetikzlibrary{patterns}
\usetikzlibrary{arrows,backgrounds,calc,chains,
	matrix,positioning,shapes,shapes.geometric,decorations,
	shapes.arrows, decorations.pathmorphing, decorations.pathreplacing, decorations.markings}

\tikzstyle{vecArrow} = [thick, decoration={markings,mark=at position
	1 with {\arrow[semithick]{open triangle 60}}},
double distance=1.4pt, shorten >= 5.5pt,
preaction = {decorate},
postaction = {draw,line width=1.4pt, white,shorten >= 4.5pt}]

\usetikzlibrary{arrows}
\usetikzlibrary{arrows.meta}
\usetikzlibrary{positioning}  
\usetikzlibrary{shadows}
\usetikzlibrary{backgrounds,calc,chains,matrix,positioning,shapes,shapes.geometric,shapes.arrows}

\tikzset{%
	font={\footnotesize},
	vertex/.style={draw,circle,inner sep=0pt,minimum width=0.5cm,minimum height=0.5cm},
	zeroterm/.style={below,inner sep=0pt,font=\tiny}
}

\usetikzlibrary{shapes.gates.logic.US}
\tikzstyle{branch}=[fill, shape=circle, minimum size=3pt, inner sep=0pt]

\usepackage{csvsimple}
\usepackage{siunitx}
\usepackage{xstring}
\newcommand{\optnum}[1]{\IfInteger{#1}{\ifnum0=0#1\relax0*\else\num{#1}\fi}{--}}
\newtheorem{myexample}{Example}

\usepackage{amsmath}

\usepackage{amssymb}
\usepackage{url}

\let\OLDthebibliography\thebibliography
\renewcommand\thebibliography[1]{
	\OLDthebibliography{#1}
	\setlength{\parskip}{0pt}
	\setlength{\itemsep}{1pt plus 0.0ex}
}

\begin{document}
\title{Design Automation for Adiabatic Circuits\vspace{.2cm}}

\author{{Alwin Zulehner$^1$\hspace{1cm}Michael P. Frank$^2$\hspace{1cm}Robert Wille$^1$} \\
{\normalsize	$^1$Institute for Integrated Circuits, Johannes Kepler University Linz, Austria} \\
{\normalsize $^2$Center for Computing Research, Sandia National Laboratories, Albuquerque, USA}\\
{\normalsize alwin.zulehner@jku.at \hspace{1cm} mpfrank@sandia.gov \hspace{1cm} robert.wille@jku.at}\\
\vspace{.4cm}	}
\date{}

\begin{abstract}Adiabatic circuits are heavily investigated since they allow for computations with an asymptotically close to zero energy dissipation per operation---serving as an alternative technology for many scenarios where energy efficiency is preferred over fast execution.
	Their concepts are motivated by the fact that the information lost from conventional circuits results in an entropy increase which causes energy dissipation. To overcome this issue, computations are performed in a (conditionally) reversible fashion
	which, additionally, have to satisfy switching rules that are different from conventional circuitry---crying out for dedicated design automation solutions.
	While previous approaches either focus on their electrical realization (resulting in small, hand-crafted circuits only) or on designing fully reversible building blocks
	(an unnecessary overhead), this work aims for providing an automatic \emph{and} dedicated design scheme that explicitly takes the recent findings in this domain into account. 
	To this end, we review the theoretical and technical background of adiabatic circuits and present  automated methods that dedicatedly realize the desired function as an adiabatic circuit. The resulting methods are further optimized---leading to an automatic and efficient design automation for this promising technology. Evaluations confirm the benefits and applicability of the proposed solution.
\end{abstract}

\maketitle

\section{Introduction}
\label{sec:intro}

As we approach the end of the semiconductor roadmap~\cite{semiconductor2015}, we are entering a regime in which fundamental thermodynamic considerations limit the sub-threshold slope, practical switching voltages, and gate energies---implying that further downscaling of device sizes and gate capacitances will soon no longer yield improvements in energy efficiency for conventional logic.
Industry's shift towards 3D geometries~\cite{semiconductor2015} will somewhat reduce parasitic energy losses in circuit structures, but once that line of improvements is played out, the only remaining approach to further increase energy efficiency will be to begin applying techniques of energy recovery. 
In this regard, resonant circuit techniques to recycle and reuse logic signal energies, rather than dissipating the entire $\frac{1}{2}CV^2$  circuit node energy on each logic-level transition, are promising. Unlike all other options, no fundamental theoretical limits on the ultimate energy efficiency of energy recovery are currently known for this direction---offering a path towards future growth of computing performance within any given energy dissipation constraints. 

But apparently the ideal of 100\% energy recovery implies that all switching activity of a device must be carried out in a manner that is asymptotically adiabatic---avoiding any abrupt loss of signal energy to heat. This motivated the consideration of \emph{adiabatic circuits} which allow for computations with an asymptotically close to zero energy dissipation (at the expense of a slower execution). 
Due to Landauer's limit~\cite{Landauer61}, this in turn implies that the computational function of the switching circuit must be \emph{logically reversible}, in the appropriately generalized sense discussed in~\cite{DBLP:conf/rc/Frank17}. Otherwise, the information lost from a conventional circuit leads to an entropy increase and, therefore, to an irreducible energy dissipation. This was recently also advocated to a larger community in~\cite{frank2017reversible} stating that the future of computing depends on reversible computations.

While these concepts have already been around for a while---general techniques for designing fully-adiabatic and reversible circuits have been introduced in the 1990's and resulted in a large body of literature (see e.g.~\cite{koller1992adiabatic,hall1992electroid,merkle1992towards,younis1993practical})---most of the adiabatic design families that have been proposed contain flaws preventing them from being truly adiabatic~\cite{DBLP:conf/csreaESA/Frank03}. In this regard, \emph{two-level adiabatic logic} (2LAL as proposed in~\cite{anantharam2004driving}) represents a very promising, \mbox{fully-adiabatic} transmission-gate logic family that relies on simple but rather efficient building blocks.
However, to realize correct adiabatic and reversible circuit designs that could truly approach arbitrarily low levels of energy dissipation
requires to satisfy certain \emph{switching rules} which differ from the design of conventional circuitry---crying out  for automated approaches for the design of such adiabatic circuits. Heading into this direction recently also gained relevance in industry---triggered e.g.~by investments of funding agencies and national departments~\cite{frank2017reversible}. Accordingly, researchers started to work towards such solutions. 

However, previously proposed approaches either focus on their electrical realization (see e.g.~\cite{younis1993practical,anantharam2004driving}) or on designing purely reversible building blocks like Toffoli gates (see e.g.~\cite{DBLP:journals/tcad/MorrisonR14,rauchenecker2017exploiting} in combination with corresponding synthesis approaches such as~\cite{zulehner2017one,ZulehnerW18Exploiting}). While the former approaches are restricted to small and \mbox{hand-crafted} circuits only, relying on purely reversible building blocks results in an unnecessarily large overhead.  
Instead, recent findings (summarized in~\cite{DBLP:conf/rc/Frank17}) show that conditional reversibility is sufficient for adiabatic circuits. But
thus far, no design automation approach for adiabatic circuits exists which exploits that in an automatic fashion.

In this work, we overcome this issue by combining expertise from both adiabatic circuits and design automation. More precisely, 
we review the theoretical and technical background of adiabatic circuits and, based on that, propose 
an automatic \emph{and} dedicated design flow for this promising technology. 
Two complementary design styles (namely retractile and fully-pipelined) are thereby considered which allow for the generation of adiabatic circuits either focusing on reducing the number of gates or keeping the number of so-called power clocks small. Furthermore, optimizations for both design styles
are proposed which utilize application-specific properties and, by this, allow e.g.~for a reduction in the number of gates by approx. 37\% and 30\% on average for the  retractile and fully-pipelined design styles, respectively.
Evaluations confirm the benefits and applicability of the proposed solution.

The remainder of this work is structured as follows:
Section~\ref{sec:background} provides a review of the theoretical and technical background of adiabatic circuits. Based on that, the proposed design flow is introduced in Section~\ref{sec:flow} followed by the descriptions of the respective mapping methods following the retractile and fully-pipelined design style in Section~\ref{sec:retractile} and Section~\ref{sec:pipelined}, respectively. Finally, a summary of the results from our evaluations is given in Section~\ref{sec:exp} and the paper is concluded in Section~\ref{sec:conclusions}.

\vfill

\section{Adiabatic Circuits}
\label{sec:background}

In this work, we consider design automation for adiabatic circuits according to the \emph{two-level adiabatic logic} (2LAL,~\cite{anantharam2004driving}) circuit family. This type of adiabatic circuit uses only two different voltage levels and heavily relies on transmission gates. 
Furthermore, a dual-rail encoding is used for the signals of the circuit, i.e.~each signal occurs in uncomplemented as well as in complemented form.\footnote{The uncomplemented form of a signal is labeled with a subscript $N$ and the complemented form is labeled with subscript $P$.}

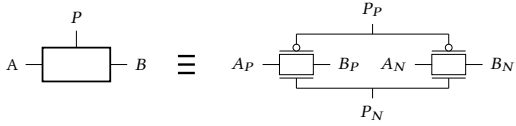
\begin{figure}\centering
	\scalebox{0.9}{
	\begin{tikzpicture}
	\draw[line width=0.75pt] (0,0) rectangle ++(1,0.5);
	
	\draw (-0.25,0.25) node[left] {A}  -- (0, 0.25);
	\draw (1,0.25) -- (1.25,0.25) node[right] {$B$};
	\draw (0.5,0.5) -- (0.5, 0.75) node[above] {$P$} ;
	
	\draw (3.25,0.25) node[left] {$A_P$} -| (3.5,0.4) -| (4,0.25) -- (4.25,0.25) node[right] {$B_P$};
	\draw (3.5,0.25) |- (4,0.1) -- (4, 0.25);
	\draw (3.5,0.05) -- (4, 0.05);
	\draw (3.5,0.45) -- (4, 0.45);	
	\draw[fill=white] (3.75,0.5) circle(0.05);
	
	\draw (3.75,0.55) -- (3.75,0.7); 
	\draw (3.75,0.05) -- (3.75,-0.1); 
	
	\draw[line width = 1pt] (2,0.125) -- (2.25,0.125);
	\draw[line width = 1pt] (2,0.25) -- (2.25,0.25);
	\draw[line width = 1pt] (2,0.375) -- (2.25,0.375);

	\draw (5.5,0.25) node[left] {$A_N$} -| (5.75,0.4) -| (6.25,0.25) -- (6.5,0.25) node[right] {$B_N$};
\draw (5.75,0.25) |- (6.25,0.1) -- (6.25, 0.25);
\draw (5.75,0.05) -- (6.25, 0.05);
\draw (5.75,0.45) -- (6.25, 0.45);	
\draw[fill=white] (6,0.5) circle(0.05);

\draw (6,0.55) -- (6,0.7); 
\draw (6,0.05) -- (6,-0.1);

\draw (6,-0.1) -- (3.75,-0.1);
\draw (4.875,-0.1) -- (4.875,-0.25) node[below] {$P_N$};

\draw (6,0.7) -- (3.75,0.7);
\draw (4.875,0.7) -- (4.875,0.85) node[above] {$P_P$};
	
	\end{tikzpicture}}

	\vspace*{-3mm}
	\caption{Transmission gate for dual-rail signals}
	\label{fig:transmission_gate}
	\vspace*{-3mm}
\end{figure}

Fig.~\ref{fig:transmission_gate} provides the notation for transmission gates: If the signal $P$ is 1 (the gate is \emph{turned on}), $A$ and $B$ are connected.\footnote{Note that logic 1 (i.e.~$X=1$) is realized by $X_N = 1$ and $X_P=0$ since a dual-rail encoding is employed.} Otherwise (the gate is \emph{turned off}), $A$ and $B$ are not connected. Since $A$ and~$B$ are both encoded in a dual-rail fashion and, thus, have an uncomplemented as well as complemented form, two transmission gates as shown in the right-hand side of Fig.~\ref{fig:transmission_gate} are required.\footnote{For sake of simplicity, we abstract the two transmission gates in the following illustrations and use the more compact form as shown in the left-hand side of Fig.~\ref{fig:transmission_gate} instead.} 
The general \emph{switching rules} for transistors in adiabatic circuits (e.g.~outlined in~\cite{anantharam2004driving,DBLP:conf/csreaESA/Frank03}) imply that   
a transmission gate shall never be turned on if $A$ and~$B$ have different values.

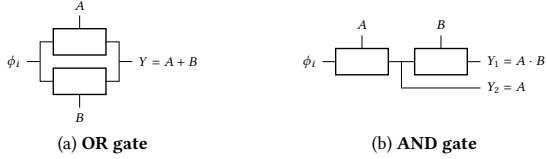
\begin{figure}
 	\begin{subfigure}[b]{0.44\linewidth}
 		\centering
\scalebox{0.7}{
 		\begin{tikzpicture}
 		\draw[line width=0.75pt] (0,0) rectangle ++(1,0.5);
 		
 		\draw[line width=0.75pt] (0,-0.75) rectangle ++(1,0.5);

 		\draw (0, 0.25) -- (-0.25,0.25) |- (0, -0.5) ;
 		\draw (1,0.25) -- (1.25,0.25) |- (1,-0.5);
 		\draw (0.5,0.5) -- (0.5, 0.75) node[above] {$A$};
 		\draw (0.5,-0.75) -- (0.5, -1) node[below] {$B$};
 		\draw (-0.25,-0.125) -- (-0.5,-0.125) node[left] {$\phi_i$};
 		\draw (1.25,-0.125) -- (1.5,-0.125) node[right] {$Y=A+B$};
 		
 		\end{tikzpicture}}
	\vspace*{-1mm}
	 		\caption{OR gate}\label{fig:gates_or}
 	\end{subfigure}	
 	\begin{subfigure}[b]{0.55\linewidth}
 		\centering
\scalebox{0.7}{
	 		\begin{tikzpicture}
 		\draw[line width=0.75pt] (0,0) rectangle ++(1,0.5);
 		
 		\draw[line width=0.75pt] (1.5,0) rectangle ++(1,0.5);
 		
 		\draw (0, 0.25) -- (-0.25,0.25) node[left] {$\phi_i$};
 		\draw (1,0.25) -- (1.5,0.25);
 		\draw (0.5,0.5) -- (0.5, 0.75) node[above] {$A$};
 		\draw (2,0.5) -- (2, 0.75) node[above] {$B$};
 		\draw (2.5,0.25) -- (2.75,0.25) node[right] {$Y_1=A\cdot B$};
 		\draw (1.25,0.25) |- (2.75,-0.25) node[right]{$Y_2=A$};
 		\draw[draw=none] (0.5, -0.625) node[below] {\phantom{$B$}};
 		\end{tikzpicture}}
	\vspace*{-1mm}
 		\caption{AND gate}
 	\end{subfigure}	
	\vspace*{-7mm}
 	\caption{Adiabatic gates}
 	\label{fig:gates}
	\vspace*{-3mm}
 \end{figure}

Besides that, so-called \emph{power clocks} (denoted $\phi_i$) are additionally utilized to realize typical functions such as OR or AND.
More precisely, the inputs of the gate control a network of transmission gates which connect the output $Y$ of the gate to one of the power clocks~$\phi_i$ in case the function to be realized evaluates to 1. 
To obey the switching rules, the output $Y$ of the gate as well as the power clock $\phi_i$ are assumed to be 0 initially. By transitioning the power clock to 1, the output of the gate is set to the desired value.
Moreover, when resetting all inputs of a gate to 0 (and, thus, disconnecting $\phi_i$ and $Y$) while $\phi_i$ is still 1, the output preserves its value (even if resetting $\phi_i$ to 0 afterwards). This allows for an inherent \emph{latching} of an output value to be used by following gates. 
An example illustrates the idea:

 \begin{myexample}
 	Fig.~\ref{fig:gates} shows the 2LAL realization of an OR gate and an AND gate. The OR gate is composed of two parallel transmission gates whose outputs are connected.
 	In case $A=1$ ($B=1$), the upper (the lower) transmission gate is turned on and connects the power clock~$\phi_i$ to the output $Y$. Consequently, $Y$ is connected to $\phi_i$ if \mbox{$A+B=1$}. Transitioning $\phi_i$ to 1 sets $Y$ to the desired value. If we now reset the inputs $A$ and $B$ to 0, the output $Y$ is latched---its value is preserved even when setting $\phi_i$ back to 0 afterwards. 	 
 	 The AND gate is realized similarly as a sequence of two transmission gates. Note that a second output $Y_2=A$ is required in this case to operate the gate in an adiabatic fashion in case $A=1$ and $B=0$ when used in a \mbox{fully-pipelined} circuit (cf.~Section~\ref{sec:pipelined}).
 \end{myexample}

Once the output of a gate is not needed anymore (e.g.~by a following gate),
an essential step for adiabatic circuits is the ability to decompute it---feeding charge back to the power clocks. 
In case that the output was not latched (i.e.~the output is still connected to the power clock), it is decomputable by simply resetting the power clock to 0 (as discussed above).
If the output is latched (i.e.~it was disconnected from the power clock by setting the inputs back to~0), the power clock has to be transitioned to 1 as well, before the inputs are applied in order to obey the switching rules.
Then, the output is decomputable by transitioning the power clock back to 0. 

\begin{myexample}\label{ex:decompute}
	Consider again the 2LAL realization of an OR gate 
	(cf.~Fig.~\ref{fig:gates_or}). Assume that the output $Y=A+B$ of the gate is latched and that all other signals are set to~0. To unlatch the output $Y$, we first have to set the power clock $\phi_i$ to 1. By this, $\phi_i$ and $Y$ have the same value if they get connected by resetting the inputs to their original value.  
	Then, $Y$ is decomputable by changing the power clock $\phi_i$ back to 0---the charge representing $Y=1$ is fed back to the power supply. 
\end{myexample}

Following this main principle allows for conducting operations with an asymptotically close to zero energy dissipation (at the expense of a slower execution since more steps have to be conducted). In fact, in contrast to conventional circuits in which energy is frequently ``grounded'', adiabatic circuits allow for feeding energy back to the clocks providing the power supply. 

However, this concept of feeding back charge to the power clocks by decomputing signals demands for a logical reversibility of the underlying computations. This is because, in order to not violate the switching rules, the original input assignments have to be applied so that signals with different values are never connected (cf.~Example~\ref{ex:decompute}). 
While in the past, a pure reversible scheme has been assumed (see e.g.~\cite{DBLP:journals/tcad/MorrisonR14,rauchenecker2017exploiting}), findings recently summarized in~\cite{DBLP:conf/rc/Frank17} showed that conditional reversibility is actually sufficient for adiabatic circuits. Again, this is illustrated by means of an example:

\begin{myexample}
	Consider again the 
	OR gate shown in Fig.~\ref{fig:gates_or}. Considering the state of the signals $A$, $B$, and $Y$, the gate describes a function $f:\mathbb{B}^3 \rightarrow \mathbb{B}^3 = (A,B,Y) \rightarrow (A,B,A+B)$. 
	This function is not reversible in general, since the initial value of $Y$ can not be computed from the output values. However, the function is conditionally reversible under the precondition that the value of $Y$ is initially set to 0, i.e.~an input combination like e.g.~$(1,0,1)$ can never occur. Conditional reversibility is a much weaker constraint than unconditional reversibility (as e.g.~considered in~\cite{DBLP:journals/tcad/MorrisonR14,rauchenecker2017exploiting})---allowing to realize adiabatic gates as e.g.~shown in Fig.~\ref{fig:gates}.
\end{myexample}

Obviously, conducting computations in such a fashion requires the corresponding circuits to be designed in a significantly different fashion than conventional circuitry. Besides the generation of a proper netlist composed of transmission gates, this additionally requires dedicated power clocks which correspondingly trigger the required operations at the correct point in time. Moreover, also the design objectives change. While the number of required (transmission) gates is still a factor (e.g.~to approximate the required area), their impact on energy consumption is smaller than for conventional circuits.
This is because energy is never grounded in adiabatic circuits but frequently  fed back to the power supply as described above. 
In contrast, the number of power clocks is much more crucial as they are the entities which actually require energy and whose waveform might be hard to generate. Besides that, more clocks usually also require longer execution times.

\section{Proposed Design Flow}\label{sec:flow}

As discussed above, previous design methods for designing adiabatic circuits (e.g.~\cite{DBLP:journals/tcad/MorrisonR14,rauchenecker2017exploiting})
assumed the requirement of full reversibility. As recently discussed in~\cite{DBLP:conf/rc/Frank17}, this leads to a significant overhead and is not necessarily needed. In fact,  conditional reversibility as reviewed above is sufficient and constitutes a much weaker constraint.
However, thus far, no design automation for this kind of adiabatic circuits exists. Also, solely employing conventional design solutions is not an option since, despite the pure functionality, a dedicated mapping and clocking scheme is required.
In this work, we present different methods which address these issues. All of them employ thereby a two-stage process.
The first step is similar to the design of conventional circuits: We realize the function to be synthesized with respect to a certain logic gate library. Afterwards, the resulting netlist is mapped to an adiabatic circuit which respectively satisfies and optimizes the rules and objectives reviewed in Section~\ref{sec:background}.

For the first part, we utilize a solution based on \emph{AND-Inverter Graphs} (AIGs~\cite{KPKG:2002}) which realize the function to be synthesized in terms of NAND gates.\footnote{Note that the design methods proposed in this work can correspondingly be adjusted to any other synthesis solution and, hence, logic gate library as well.}
AIGs allow for a graph-based representation of Boolean functions. The graph has one root node for each output of the function. The inputs of the function are provided as terminals. The intermediate nodes of an AIG represent an AND operation and, thus, have two successors each. To gain universality, the inputs of the AND operation can be inverted. This is denoted by black circles on the respective edges.
Equal nodes occur frequently and can be shared---allowing for a compact representation of the function to be realized. 

\begin{figure}
	\begin{subfigure}[b]{0.49\linewidth}
		\centering
	\scalebox{0.75}{
		\centering
		\begin{tikzpicture}[terminal/.style={draw,rectangle,inner sep=2pt}]
		\matrix[matrix of nodes,ampersand replacement=\&,every node/.style={vertex},column sep={1.5cm,between origins},row sep={0.9cm,between origins}] (qmdd) {
			\node[regular polygon,regular polygon sides=4, inner sep=1pt] (y1) {$y_1$}; \& \node[regular polygon,regular polygon sides=4, inner sep=1pt] (y0) {$y_0$}; \& \\
			\node(n1a) {$\wedge_6$}; \& \node(n1b) {$\wedge_7$}; \& \\
			\& \node (n2a) {$\wedge_4$}; \& \node (n2b) {$\wedge_5$}; \\
			\node(n3a) {$\wedge_2$}; \& \node(n3b) {$\wedge_3$}; \& \\
			\& \node(n4) {$\wedge_1$}; \& \& \\
			\node[regular polygon,regular polygon sides=3,draw,inner sep=0pt] (x2) {$x_2$}; \& 		\node[regular polygon,regular polygon sides=3,draw,inner sep=0pt] (x1) {$x_1$}; \& 		\node[regular polygon,regular polygon sides=3,draw,inner sep=0pt] (x0) {$x_0$}; \\
		};
		
		\draw (x2.north) -- (n3a);
		\draw (x2.north) -- (n4) node[midway, circle, inner sep=1pt, fill] {};
		\draw (x1.north) -- (n3a);
		\draw (x1.north) -- (n4) node[midway, circle, inner sep=1pt, fill] {};
		\draw (x1.north) -- (n2b);
		\draw (x0.north) -- (n3b) node[midway, circle, inner sep=1pt, fill] {};
		\draw (x0.north) -- (n2b);
		\draw (n4) -- (n3b) node[midway, circle, inner sep=1pt, fill] {};
		\draw (n3a) -- (n1a) node[midway, circle, inner sep=1pt, fill] {};
		\draw (n3a) -- (n2a) node[midway, circle, inner sep=1pt, fill] {};
		\draw (n3b) -- (n1a) node[midway, circle, inner sep=1pt, fill] {};
		\draw (n3b) -- (n2a);
		\draw (n2a) -- (n1b) node[midway, circle, inner sep=1pt, fill] {};
		\draw (n2b) -- (n1b) node[midway, circle, inner sep=1pt, fill] {};
		\draw (n1a) -- (y1);
		\draw (n1b) -- (y0) node[midway, circle, inner sep=1pt, fill] {};
				
		\end{tikzpicture}}
	
		\caption{AND-Inverter graph}
		\label{fig:aig}
	\end{subfigure}
	\begin{subfigure}[b]{0.49\linewidth}
		\centering
		\scalebox{0.75}{
			\centering
		\begin{tikzpicture}[terminal/.style={draw,rectangle,inner sep=2pt}]
		\matrix[matrix of nodes,ampersand replacement=\&,every node/.style={vertex},column sep={1.5cm,between origins},row sep={0.9cm,between origins}] (qmdd) {
			\node[regular polygon,regular polygon sides=4, inner sep=1pt] (y1) {$y_1$}; \& \node[regular polygon,regular polygon sides=4, inner sep=1pt] (y0) {$y_0$}; \& \\
			\node(n1a) {$\vee_6$}; \& \node(n1b) {$\vee_7$}; \& \\
			\& \node (n2a) {$\vee_4$}; \& \node (n2b) {$\vee_5$}; \\
			\node(n3a) {$\vee_2$}; \& \node(n3b) {$\vee_3$}; \& \\
			\& \node(n4) {$\vee_1$}; \& \& \\
			\node[regular polygon,regular polygon sides=3,draw,inner sep=0pt] (x2) {$x_2$}; \& 		\node[regular polygon,regular polygon sides=3,draw,inner sep=0pt] (x1) {$x_1$}; \& 		\node[regular polygon,regular polygon sides=3,draw,inner sep=0pt] (x0) {$x_0$}; \\
		};
		
		\draw (x2.north) -- (n3a) node[midway, circle, inner sep=1pt, fill] {};
		\draw (x2.north) -- (n4);
		\draw (x1.north) -- (n3a) node[midway, circle, inner sep=1pt, fill] {};
		\draw (x1.north) -- (n4);
		\draw (x1.north) -- (n2b) node[midway, circle, inner sep=1pt, fill] {};
		\draw (x0.north) -- (n3b);
		\draw (x0.north) -- (n2b) node[midway, circle, inner sep=1pt, fill] {};
		\draw (n4) -- (n3b) node[midway, circle, inner sep=1pt, fill] {};
		\draw (n3a) -- (n1a) node[midway, circle, inner sep=1pt, fill] {};
		\draw (n3a) -- (n2a) node[midway, circle, inner sep=1pt, fill] {};
		\draw (n3b) -- (n1a) node[midway, circle, inner sep=1pt, fill] {};
		\draw (n3b) -- (n2a);
		\draw (n2a) -- (n1b) node[midway, circle, inner sep=1pt, fill] {};
		\draw (n2b) -- (n1b) node[midway, circle, inner sep=1pt, fill] {};
		\draw (n1a) -- (y1) node[midway, circle, inner sep=1pt, fill] {};
		\draw (n1b) -- (y0);

		\end{tikzpicture}}
			\caption{OR-Inverter graph}
		\label{fig:oig}
	\end{subfigure}
	\vspace*{-2mm}
	\caption{Graph representations for Boolean functions}
	\vspace*{-2mm}
\end{figure}
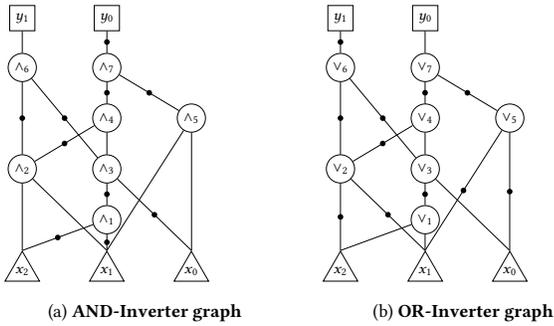

\begin{myexample}
	Fig.~\ref{fig:aig} shows the AIG of a 3-input 2-output Boolean function with inputs $x_2$, $x_1$, and $x_0$ as well as outputs 
	$y_1$ and $y_0$ which represent $y_1 = \overline{x}_2\overline{x}_1 + \overline{x}_2x_0 + \overline{x}_1x_0$ and $y_0 = \overline{x}_2x_1+x_1x_0+x_2\overline{x}_1\overline{x}_0$ in terms of an AIG and, hence, NAND operations. 
\end{myexample}

How to determine and optimize an AIG  (e.g.~minimizing its number of nodes/gates) has intensely been considered in the literature (see e.g.~\cite{DBLP:conf/dac/MishchenkoCB06}) and, hence, is not covered further in the following. Instead, we focus on the second step, i.e.~how to map the resulting NAND netlist to an adiabatic circuit, i.e.~a network of transmission gates and the corresponding power clocks.
To this end, we translate the AIG into an \emph{OR-Inverter graph} (OIG) so that a NOR gate netlist results. An OIG can easily be derived from an AIG by simply applying De Morgan's laws, i.e.~by relabeling the inner nodes from AND to OR and inverting the polarity of the edges to the terminals and the edges to the root nodes (cf.~\ref{fig:oig}). 

Now, the nodes of an OIG can directly be  mapped to the adiabatic OR gates introduced before in Fig~\ref{fig:gates_or}.
However, it remains open and non-trivial how to connect these gates to the power clocks and how to generate a corresponding waveform of these clocks (again, following the switching rules and optimization objectives reviewed in Section~\ref{sec:background}).
To this end, two (complementary) design styles are considered: \emph{retractile} circuits (cf.~\cite{hall1992electroid}) as well as \mbox{\emph{fully-pipelined}} circuits (cf.~\cite{younis1993practical,anantharam2004driving,DBLP:conf/rc/Frank17}). 
Note that for both design styles the conditional reversibility is inherently satisfied by preserving the inputs of the signals throughout the whole computation and by assuming that all additional (intermediate) signals are initially set to 0. 
In the following sections, we discuss advantages and disadvantages of both design styles and present according (automatic) mapping schemes. 
More precisely, for each design style we first describe a straightforward mapping scheme (conveying the main idea of the design style) followed by an advanced mapping scheme (which results in a significantly smaller number of gates as well as, in case of retractile circuits, to a smaller number of power clocks).
These considerations eventually motivate the implementation of different methods for design automation of adiabatic circuits whose performance is eventually discussed in Section~\ref{sec:exp}.

\section{Retractile Circuits}
\label{sec:retractile}

\subsection{Straightforward Solution}
\label{sec:retractile_sf}

The straightforward mapping for retractile circuits is similar to conventional circuitry, where an AIG or OIG is directly mapped to the target technology. In fact, we can realize each node of the OIG with an OR gate and negations with inverters. 
Moreover, in case of adiabatic circuits, the inverters come ``for free'' since we are operating on dual-rail signals and, hence, an inverted input can easily be realized with no further hardware by swapping the rails of the signal.

\begin{figure}
\begin{subfigure}{0.58\linewidth}
	\centering
	\scalebox{0.7}{
\begin{tikzpicture}
\node (x2) at (0, 2) {$x_2$};
\node (x1) at (0, 1) {$x_1$};
\node (x0) at (0, 0) {$x_0$};

\node[or gate US, draw, anchor=input 1] at ($(x2) + (2.5, 0)$) (or2) {$g_2$};
\node[or gate US, draw, anchor=input 2] at ($(x1) + (1, 0)$) (or1) {$g_1$};
\node[or gate US, draw, anchor=input 1] at (2.5,0 |- or1.output) (or3) {$g_3$};
\node[or gate US, draw, anchor=input 2] at (4,0 |- or3.output) (or4) {$g_4$};
\node[or gate US, draw, anchor=input 2] at ($(x0) + (4, 0)$) (or5) {$g_5$};
\node[or gate US, draw, anchor=input 1] at (5.5,0 |- or2.output) (or6) {$g_6$};

\coordinate (pos7) at ($(or4.output)!.5!(or5.output)$);
\node[or gate US, draw, anchor=output] at (or6.output |- pos7) (or7) {$g_7$};

\draw (x2) -- (or2.input 1);
\draw (x1) -- (or1.input 2);
\draw (or1.output) -- (or3.input 1);
\draw (x0) -- (or5.input 2);
\draw (or3.output) -- (or4.input 2);
\draw (or2.output) -- (or6.input 1);

\draw ($(x2)+(0.75,0)$) node[branch] {} |- ($(or1.input 1)$);
\draw ($(x1)+(0.5,0)$) node[branch] {} |- ($(or2.input 2)$);
\draw ($(x0)+(2,0)$) node[branch] {} |- ($(or3.input 2)$);
\draw ($(or2.output)+(0.5,0)$) node[branch] {} |- ($(or4.input 1)$);
\draw ($(or3.output)+(0.25,0)$) node[branch] {} |- ($(or6.input 2)$);
\draw ($(x1)+(0.5,0)$) |- ($(or5.input 1)$);
\draw (or5.output) -- ++ (0.5,0) |- ($(or7.input 2)$);
\draw (or4.output) -- ++ (0.5,0) |- ($(or7.input 1)$);
\draw (or6.output) -- ++ (0.25,0) node[anchor=west] {$y_1$};
\draw (or7.output) -- ++ (0.25,0) node[anchor=west] {$y_0$};

\draw[line width=0.4, fill=white] ($(or2.input 1)-(0.07,0)$) circle (0.07);
\draw[line width=0.4, fill=white] ($(or2.input 2)-(0.07,0)$) circle (0.07);
\draw[line width=0.4, fill=white] ($(or3.input 1)-(0.07,0)$) circle (0.07);
\draw[line width=0.4, fill=white] ($(or4.input 1)-(0.07,0)$) circle (0.07);
\draw[line width=0.4, fill=white] ($(or5.input 1)-(0.07,0)$) circle (0.07);
\draw[line width=0.4, fill=white] ($(or5.input 2)-(0.07,0)$) circle (0.07);
\draw[line width=0.4, fill=white] ($(or6.input 1)-(0.07,0)$) circle (0.07);
\draw[line width=0.4, fill=white] ($(or6.input 2)-(0.07,0)$) circle (0.07);
\draw[line width=0.4, fill=white] ($(or7.input 1)-(0.07,0)$) circle (0.07);
\draw[line width=0.4, fill=white] ($(or7.input 2)-(0.07,0)$) circle (0.07);
\draw[line width=0.4, fill=white] ($(or5.input 1)-(0.07,0)$) circle (0.07);
\draw[line width=0.4, fill=white] ($(or6.output)+(0.07,0)$) circle (0.07);

\draw[dashed] (1.75,-0.25) -- (1.75,2.5);
\node[above] at (1,2.25) {$s_0$};
\draw[dashed] (3.25,-0.25) -- (3.25,2.5);
\node[above] at (2.5,2.25) {$s_1$};
\draw[dashed] (4.75,-0.25) -- (4.75,2.5);
\node[above] at (4,2.25) {$s_2$};
\node[above] at (5.5,2.25) {$s_3$};

\end{tikzpicture}}

\caption{Circuit}
\label{fig:circuit_retractile_gates}
\end{subfigure}
	\begin{subfigure}{0.39\linewidth}
	\centering

\scalebox{0.6}{
	\begin{tikzpicture}
	\draw[->](0,0) -- (5.0, 0) node[right] {$t$};
	\draw[-](0, 0) -- (0, 2.5) node[above] {};
	
	\draw[domain=0:.5, smooth, variable=\x, red]
	plot ({\x}, {(\x)});
	\draw[domain=.5:4, smooth, variable=\x, red]
	plot ({\x}, {.5});
	\draw[domain=4:4.5, smooth, variable=\x, red]
	plot ({\x}, {-\x+4.5});
	
	\draw[domain=0:0.5, smooth, variable=\x, red]
	plot ({\x}, {0.6});
	\draw[domain=0.5:1, smooth, variable=\x, red]
	plot ({\x}, {0.1+(\x)});
	\draw[domain=1:3.5, smooth, variable=\x, red]
	plot ({\x}, {1.1});
	\draw[domain=3.5:4, smooth, variable=\x, red]
	plot ({\x}, {-\x+4.6});
	\draw[domain=4:4.5, smooth, variable=\x, red]
	plot ({\x}, {0.6});
	
	\draw[domain=0:1, smooth, variable=\x, red]
	plot ({\x}, {1.2});
	\draw[domain=1:1.5, smooth, variable=\x, red]
	plot ({\x}, {0.2+(\x)});
	\draw[domain=1.5:3, smooth, variable=\x, red]
	plot ({\x}, {1.7});
	\draw[domain=3:3.5, smooth, variable=\x, red]
	plot ({\x}, {-\x+4.7});
	\draw[domain=3.5:4.5, smooth, variable=\x, red]
	plot ({\x}, {1.2});
	
	\draw[domain=0:1.5, smooth, variable=\x, red]
	plot ({\x}, {1.8});
	\draw[domain=1.5:2, smooth, variable=\x, red]
	plot ({\x}, {0.3+(\x)});
	\draw[domain=2:2.5, smooth, variable=\x, red]
	plot ({\x}, {2.3});
	\draw[domain=2.5:3, smooth, variable=\x, red]
	plot ({\x}, {-\x+4.8});
	\draw[domain=3:4.5, smooth, variable=\x, red]
	plot ({\x}, {1.8});
	
	\node[anchor = center] at (-0.25,0.25) {$\phi_0$};
	\node[anchor = center] at (-0.25,0.85) {$\phi_1$};
	\node[anchor = center] at (-0.25,1.45) {$\phi_2$};
	\node[anchor = center] at (-0.25,2.05) {$\phi_3$};
	
	\draw (0,-0.1) node[below] {0};
	\foreach \x in {1,...,9} {
		\draw (0.5*\x,0.1) -- (0.5*\x,-0.1) node[below]{\x};
	}
	
	\end{tikzpicture}
}

	\caption{Power clocks}
	\label{fig:clocks}
\end{subfigure}
	\vspace*{-2mm}
	\caption{Synthesized retractile circuit}
\label{fig:circuit_retractile}	
	\vspace*{-2mm}
\end{figure}
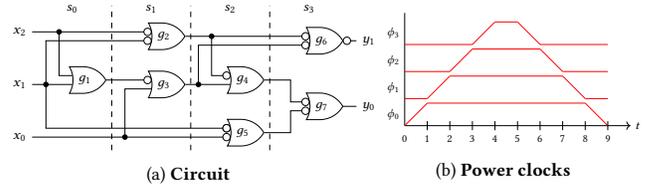

\begin{myexample}
	Consider again the OIG depicted in Fig.~\ref{fig:oig}. Mapping the OIG to conventional gates results in the circuit shown in Fig.~\ref{fig:circuit_retractile_gates}. 
	Doing this mapping for adiabatic circuits following the retractile design style, each OR-gate is realized with two transmission gates as discussed in Section~\ref{sec:background}. 
\end{myexample}

To operate the circuit in an adiabatic fashion, all intermediate signals are first initialized with~0. Furthermore,
each stage $s_i$ ($0 \le i < N$) of the circuit with depth $N$
has an associated dual-rail encoded clock $\phi_i$---allowing to compute the individual stages sequentially. Then, the computations are started by transitioning the $0^{th}$ clock from 0 to 1---triggering the desired operations of the first stage. 
Once stable, the operations of the next stages are sequentially triggered. 
To allow for decomputing the intermediate results, the clocks transition back to 0 in reverse order, i.e.~first the $N-1^{th}$ clock is set back to~0, then the other ones. 
This way, all intermediate results are decomputed and restored back to 0. Overall, this requires $2N+1$ time steps
for a single computation (assuming one additional time step is required to process the outputs of the circuit). During these time steps, the inputs have to remain constant---yielding a rather low throughput.

\addtocounter{myexample}{-1}
\begin{myexample}[continued]
	Since the resulting circuit has four stages (the OIG has a depth of 4), we need four different clocks (eight if we take the dual-rail encoding into account). The waveforms of these clocks are shown in Fig.~\ref{fig:clocks}. Overall, this causes that a single computation of this circuit requires 9 timesteps.
\end{myexample}

\subsection{Advanced Solution}
\label{sec:retractile_opt}

\begin{figure}
	\centering
	\scalebox{0.7}{
\begin{tikzpicture}

\node[or gate US, draw] (or1) {};
\node[or gate US, draw, below = 0.25cm of or1] (or2) {}; 
\coordinate (pos7) at ($(or1.output)!.5!(or2.output)$);
\node[or gate US, draw] at (1.15,0 |- pos7) (or3) {};

\node[above=-0.1cm of or1,xshift=0.4cm] {fanout $=$ 1};
\node[below=-0.1cm of or2,xshift=0.4cm] {fanout $=$ 1};

\draw (or1.output) -- ++ (0.25,0) |- ($(or3.input 1)$);
\draw (or2.output) -- ++ (0.25,0) |- ($(or3.input 2)$);
\draw (or1.input 1) -- ++ (-0.25,0);
\draw (or1.input 2) -- ++ (-0.25,0);
\draw (or2.input 1) -- ++ (-0.25,0);
\draw (or2.input 2) -- ++ (-0.25,0);
\draw (or3.output) -- ++ (0.25,0);
\draw[dashed] ($(or1.input 1)+(-0.15,0.5)$) |- ($(or2.input 2)+(1.95,-0.5)$) |- ($(or1.input 1)+(-0.15,0.5)$);

\node[or gate US, draw, right=2.9cm of or1] (or4) {};
\node[or gate US, draw, below = 0.25cm of or4] (or5) {}; 
\coordinate (pos8) at ($(or4.output)!.5!(or5.output)$);

\draw (or4.output) -- ++ (0.25,0) |- ($(pos8)+(0.55,0)$);
\draw (or5.output) -- ++ (0.25,0) |- ($(pos8)+(0.55,0)$) node[midway, branch] {};
\draw (or4.input 1) -- ++ (-0.25,0);
\draw (or4.input 2) -- ++ (-0.25,0);
\draw (or5.input 1) -- ++ (-0.25,0);
\draw (or5.input 2) -- ++ (-0.25,0);
\draw[dashed] ($(or4.input 1)+(-0.15,0.5)$) |- ($(or5.input 2)+(1.1,-0.5)$) |- ($(or4.input 1)+(-0.15,0.5)$);

\path[draw=black,solid,line width=2mm,fill=black,
preaction={-triangle 90,thin,draw,shorten >=-1mm}
]($(or3.output)+(0.4,0)$)-- ++ (0.8,0) node[above, midway]{Rule 1};

\node[or gate US, draw, right = 5.5 cm of or1] (or6) {};
\node[or gate US, draw, below = 0.25cm of or6] (or7) {}; 
\coordinate (pos1) at ($(or6.output)!.5!(or7.output)$);
\node[or gate US, draw] at (7.3,0 |- pos1) (or8) {};

\node[above=-0.1cm of or6,xshift=0.4cm] {fanout $\neq$ 1};
\node[below=-0.1cm of or7,xshift=0.4cm] {fanout $=$ 1};

\draw (or6.output) -- ++ (0.25,0) |- ($(or8.input 1)$);
\draw (or7.output) -- ++ (0.25,0) |- ($(or8.input 2)$);
\draw (or6.input 1) -- ++ (-0.25,0);
\draw (or6.input 2) -- ++ (-0.25,0);
\draw (or7.input 1) -- ++ (-0.25,0);
\draw (or7.input 2) -- ++ (-0.25,0);
\draw (or8.output) -- ++ (0.25,0);
\draw ($(or6.output)+(0.25,0)$) node[branch] {} |- ($(or6.output)+(0.7,0.75)$);
\draw[dashed] ($(or6.input 1)+(-0.15,0.5)$) |- ($(or7.input 2)+(1.95,-0.5)$) |- ($(or6.input 1)+(-0.15,0.5)$);

\node[or gate US, draw, right = 2.9cm of or6] (or9) {};
\node[or gate US, draw=none, below = 0.25cm of or9] (or10) {}; 
\coordinate (pos11) at ($(or9.output)!.5!(or10.output)$);
\node[or gate US, draw] at ($(or10)+(0.9,0)$) (or11) {};
\node[buffer gate US, draw, anchor=output] at (or11.output |- or9.output) (buf) {};

\draw (or9.output) -- (buf.input);
\draw ($(or10.input 1)+(-0.25,0)$) -- (or11.input 1);
\draw ($(or10.input 2)+(-0.25,0)$) -- (or11.input 2);
\draw (or9.input 1) -- ++ (-0.25,0);
\draw (or9.input 2) -- ++ (-0.25,0);

\coordinate (pos2) at ($(or11.output)!.5!(buf.output)$);
\draw (buf.output) -- ++ (0.2,0) |- ($(pos2)+(0.55,0)$) node[midway, branch] {};
\draw (or11.output) -- ++ (0.2,0) |- ($(pos2)+(0.55,0)$);

\draw ($(or9.output)+(0.2,0)$) node[branch] {} |- ($(or9.output)+(0.7,0.75)$);
\draw[dashed] ($(or9.input 1)+(-0.15,0.5)$) |- ($(or10.input 2)+(1.95,-0.5)$) |- ($(or9.input 1)+(-0.15,0.5)$);

\path[draw=black,solid,line width=2mm,fill=black,
preaction={-triangle 90,thin,draw,shorten >=-1mm}
]($(or8.output)+(0.4,0)$)-- ++ (0.8,0) node[above, midway]{Rule 2};

\end{tikzpicture}}
	\caption{Rules for optimization}
	\label{fig:retractile_opt_rules}
\end{figure}
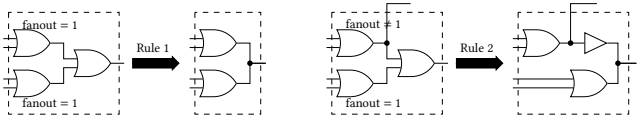

The straightforward mapping described above can significantly be optimized  to reduce the number of required transmission gates and power clocks. 
The optimized mapping scheme is motivated by an analysis of the realization of an OR gate, which is composed of two parallel buffers (i.e.~a transmission gate), whose outputs are connected (cf.~Fig.~\ref{fig:gates_or}).
Consequently, an OR gate with multiple inputs can be generated by adding further buffers in parallel. 
This way, each OIG node, whose children both have a fanout of 1 (and, thus, represents a 4-inputs OR gate) can be realized in a single stage of the circuit composed of two 2-input OR gates whose outputs are connected. 
A similar optimization can be performed for OIG nodes, where only one of the children has a fanout of 1. Here, one buffer is required for the child which has a fanout larger than one (in order to avoid sneak-paths). Additionally, the gate representing the child with fanout 1 has to be lifted to the next stage of the circuit since both, the buffer as well as the gate, have to be operated by the same power clock 
to allow for an adiabatic computation. The optimization rules are shown in Fig.~\ref{fig:retractile_opt_rules} and denoted \emph{Rule 1} and \emph{Rule 2} in the following.

Note that one has to be careful when applying the rules if the corresponding input of the gate is inverted. In this case, the inversion has to be pushed towards the inputs. This is possible by applying De Morgan's law ($\overline{a+b} = \overline{a}\cdot\overline{b}$). Consequently, we have to invert the inputs on this level and exchange the OR gate with an AND gate.\footnote{Note that this is also possible if there are several subsequent nodes for which the rules can be applied.}

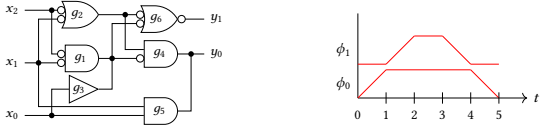
\begin{figure}[t]
\begin{minipage}[c]{0.5\linewidth}
	\centering
\scalebox{0.7}{\begin{tikzpicture}
\node (x2) at (0, 2) {$x_2$};
\node (x1) at (0, 1) {$x_1$};
\node (x0) at (0, 0) {$x_0$};

\node[or gate US, draw, anchor=input 1] at ($(x2) + (1, 0)$) (or2) {$g_2$};
\node[and gate US, draw, anchor=input 2] at ($(x1) + (1, 0)$) (or1) {$g_1$};

\node[buffer gate US, draw, anchor=output] at (or1.output |- 0,0.5) (buf) {$g_3$};
\node[and gate US, draw, anchor=input 2] at (2.5,0 |- or1.output) (or4) {$g_4$};
\node[and gate US, draw, anchor=input 2] at ($(x0) + (2.5, 0)$) (or5) {$g_5$};
\node[or gate US, draw, anchor=input 1] at (2.5,0 |- or2.output) (or6) {$g_6$};

\coordinate (pos7) at ($(or4.output)!.5!(or5.output)$);

\draw (x2) -- (or2.input 1);
\draw (x1) -- (or1.input 2);
\draw (x0) -- (or5.input 2);
\draw (or1.output) -- (or4.input 2);
\draw (or2.output) -- (or6.input 1);

\draw ($(x2)+(0.75,0)$) node[branch] {} |- ($(or1.input 1)$);
\draw ($(x1)+(0.5,0)$) node[branch] {} |- ($(or2.input 2)$);
\draw ($(x0)+(0.75,0)$) node[branch] {} |- ($(buf.input)$);
\draw ($(buf.output)$) -| ($(or1.output)+(0.25,0)$) node[branch] {};
\draw ($(or2.output)+(0.5,0)$) node[branch] {} |- ($(or4.input 1)$);
\draw ($(or1.output)+(0.25,0)$) |- ($(or6.input 2)$);
\draw ($(x1)+(0.5,0)$) |- ($(or5.input 1)$);
\draw (or5.output) -| ($(or4.output)+(0.25,0)$) node[branch] {};
\draw (or4.output) -- ++ (0.5,0) node[anchor=west] {$y_0$}; 
\draw (or6.output) -- ++ (0.5,0) node[anchor=west] {$y_1$};

\draw[line width=0.4, fill=white] ($(or1.input 1)-(0.07,0)$) circle (0.07);
\draw[line width=0.4, fill=white] ($(or1.input 2)-(0.07,0)$) circle (0.07);

\draw[line width=0.4, fill=white] ($(or2.input 1)-(0.07,0)$) circle (0.07);
\draw[line width=0.4, fill=white] ($(or2.input 2)-(0.07,0)$) circle (0.07);
\draw[line width=0.4, fill=white] ($(or4.input 2)-(0.07,0)$) circle (0.07);
\draw[line width=0.4, fill=white] ($(or6.input 1)-(0.07,0)$) circle (0.07);
\draw[line width=0.4, fill=white] ($(or6.input 2)-(0.07,0)$) circle (0.07);
\draw[line width=0.4, fill=white] ($(or6.output)+(0.07,0)$) circle (0.07);
\end{tikzpicture}}
\end{minipage}
\begin{minipage}[c]{0.49\linewidth}
	\centering
\scalebox{0.75}{
\begin{tikzpicture}
\draw[->](0,0) -- (3.0, 0) node[right] {$t$};
\draw[-](0, 0) -- (0, 1.5) node[above] {};

\draw[domain=0:.5, smooth, variable=\x, red]
plot ({\x}, {(\x)});
\draw[domain=.5:2, smooth, variable=\x, red]
plot ({\x}, {.5});
\draw[domain=2:2.5, smooth, variable=\x, red]
plot ({\x}, {-\x+2.5});

\draw[domain=0:0.5, smooth, variable=\x, red]
plot ({\x}, {0.6});
\draw[domain=0.5:1, smooth, variable=\x, red]
plot ({\x}, {0.1+(\x)});
\draw[domain=1:1.5, smooth, variable=\x, red]
plot ({\x}, {1.1});
\draw[domain=1.5:2, smooth, variable=\x, red]
plot ({\x}, {-\x+2.6});
\draw[domain=2:2.5, smooth, variable=\x, red]
plot ({\x}, {0.6});

\node[anchor = center] at (-0.25,0.25) {$\phi_0$};
\node[anchor = center] at (-0.25,0.85) {$\phi_1$};

\draw (0,-0.1) node[below] {0};
\foreach \x in {1,...,5} {
	\draw (0.5*\x,0.1) -- (0.5*\x,-0.1) node[below]{\x};
}

\end{tikzpicture}}
\end{minipage}

	\vspace*{-2mm}
	\caption{Optimized retractile circuit}
	\label{fig:circuit_retractile_opt}
	\vspace*{-2mm}
\end{figure}

\begin{myexample}
	Consider again the circuit shown in Fig.~\ref{fig:circuit_retractile}. The children of gate $g_7$ (i.e.~$g_4$ and $g_5$) both have a fanout of 1. Consequently, we can apply \emph{Rule 1} to remove $g_7$. Furthermore, one child of gate $g_3$ has a fanout of 1 (i.e.~the input $x_0$). Consequently, we can apply \emph{Rule 2} for gate $g_3$. The resulting (optimized) circuit is shown in Fig.~\ref{fig:circuit_retractile_opt}. Since both inputs of $g_7$ and one input of $g_3$ are inverted, we have to apply De Morgan's law. Consequently, the gates $g_1$, $g_4$, and $g_5$ are transformed into an AND-gate. The resulting circuit only requires 11 transmission gates and has only two stages (and, thus, suddenly requires only two different dual-rail encoded clocks).
\end{myexample}

\section{Fully-Pipelined Circuits}
\label{sec:pipelined}

The main disadvantages of the retractile circuits considered in Section~\ref{sec:retractile} are that many different power clocks are required (one for each stage) and that a computation can be conducted only every $2N+1$ time steps---resulting in a rather low throughput. 
These issues can be avoided by using fully-pipelined circuits. 
In conventional design, this would require a register after each stage of the circuit. For the adiabatic circuits considered here, however, this is not necessary, because the gates inherently allow for latching their output (cf.~Section~\ref{sec:background}).
In fact, we only have to compute the outputs of a stage $s_i$ while decomputing the signals of stage $s_{i-1}$ (i.e.~resetting them back to 0). This way, only two different power clocks (four if we take the dual-rail encoding into account) are required  (independent from the circuit depth) and computations can be conducted in a pipelined fashion (leading to a much higher throughput).

To realize this, however, 
the functions computed in the individual stages have to be (conditionally) reversible. This can easily be achieved by forwarding all the input signals of stage $s_{i-1}$ to the stage $s_i$ by using buffers.
The following example illustrates the idea of such buffers. 

\begin{figure}
	\centering
	\begin{subfigure}[b]{.43\linewidth}
	\centering
\scalebox{0.7}{
	\begin{tikzpicture}
	
	\draw (0.25,-1.25) node[below] {$\phi_0$}  |- (0.75, 0.5);
	\draw (0.25,0.5) -- (-0.25,0.5) node[left] {$x_{t-1}$};
	\draw (1.,1.25) node[above] {$\phi_1$}  |- (0.5, -0.5);
	\draw (1,-0.5) -- (1.5,-0.5) node[right] {$x_{t}$};
	
	\draw[line width=0.75pt, fill=white] (0,0) rectangle ++(0.5,-1);
	\draw[line width=0.75pt, fill=white] (0.75,0) rectangle ++(0.5,1); 
	
	\end{tikzpicture}}
		\captionof{figure}{Transmission gates}
	\label{fig:buffer}
\end{subfigure}
	\begin{subfigure}[b]{0.54\linewidth}
		\centering
\scalebox{0.75}{
			\begin{tikzpicture}
		\draw[->](0,0) -- (2.5, 0) node[right] {$t$};
		\draw[-](0, 0) -- (0, 1.5) node[above] {};
		
		\draw (0,-0.1) node[below] {0};
		\foreach \x in {1,...,4} {
			\draw (0.5*\x,0.1) -- (0.5*\x,-0.1) node[below]{\x};
		}
				
		\draw[domain=0:0.5, smooth, variable=\x, red]
		plot ({\x}, {0.6});
		\draw[domain=0.5:1, smooth, variable=\x, red]
		plot ({\x}, {.1+\x)});
		\draw[domain=1:1.5, smooth, variable=\x, red]
		plot ({\x}, {1.1});
		\draw[domain=1.5:2, smooth, variable=\x, red]
		plot ({\x}, {-\x+2.6});
		
		\draw[domain=0:0.5, smooth, variable=\x, red]
		plot ({\x}, {\x});
		\draw[domain=0.5:1, smooth, variable=\x, red]
		plot ({\x}, {.5});
		\draw[domain=1:1.5, smooth, variable=\x, red]
		plot ({\x}, {1.5-\x});
		\draw[domain=1.5:2, smooth, variable=\x, red]
		plot ({\x}, {0});
		
		\node[anchor = east] at (-0.0,0.85) {$\phi_1$};
		\node[anchor = east] at (-0.0,0.25) {$\phi_0$};
		\end{tikzpicture}}	
		\captionof{figure}{Clocks}
		\label{fig:clocks_fully_pipelined}
	\end{subfigure}
	\vspace*{-3mm}
	\caption{Buffer element for fully-pipelined circuits}
	\vspace*{-3mm}
\end{figure}
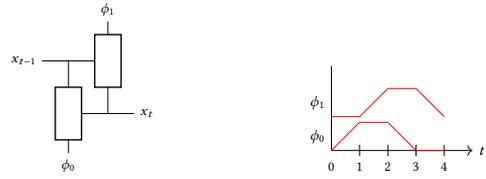

\begin{myexample}
	Fig.~\ref{fig:buffer} shows the structure of a buffer that sets \mbox{$x_{t} = x_{t-1}$} while decomputing $x_{t-1}$ (i.e.~while resetting~$x_{t-1}$ back to 0). Initially, both clocks $\phi_0$ and $\phi_1$ as well as $x_{t}$ are set to~0.
	If $x_{t-1}=1$, the transmission gate on the right connects $\phi_1$ with $x_{t}$. In the first time step, $\phi_0$ transitions to 1 (c.f.~Fig.~\ref{fig:clocks_fully_pipelined}). Afterwards, $\phi_1$ transitions to 1, setting $x_{t} = x_{t-1}$. If $x_{t}=x_{t-1}=1$, the transmission gate on the left hand side in Fig.~\ref{fig:buffer} connects $\phi_0$ with $x_{t-1}$.
	This does not violate the switching rules discussed in Section~\ref{sec:background} since $\phi_0$ is also 1. In the next time step, $\phi_0$ transitions back to 0---decomputing $x_{t-1}$ and, thus, disconnecting $\phi_i$ and $x_{t}$. Consequently, the output $x_{t}$ remains at its voltage level when eventually transitioning $\phi_1$ back to 0---the output is latched.
\end{myexample}

To allow for inverted inputs of gates, a quad-rail encoding is required for the signals to properly decompute the inputs~\cite{anantharam2004driving}. Here, each signal $X$ is represented by two dual-rail signals (one for $X=1$ and one for $X=0$). Initially, both dual-rail signals are set to 0. This again allows to realize inverters without any transmission gates---just swapping the two dual-rail signals $X=1$ and $X=0$. 
In the following we again abstract this fact when illustrating the required transmission gates.

\subsection{Straightforward Solution}
\label{sec:pipelined_sf}

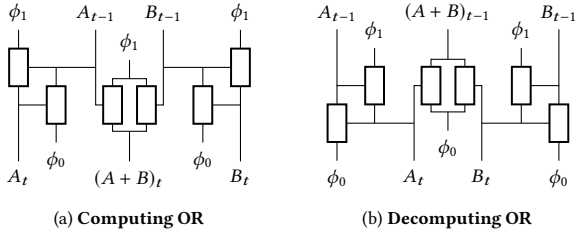
\begin{figure}
	
	\centering
	\begin{subfigure}[b]{0.49\linewidth}
	\centering
	\begin{tikzpicture}
	\draw[line width=0.75pt] (0.1,0) rectangle ++(0.25,0.5);
	\draw[line width=0.75pt] (0.85,0) rectangle ++(0.25,0.5); 
	\draw[line width=0.75pt] (1.35,0.5) rectangle ++(0.25,0.5); 
	
	\draw[line width=0.75pt] (-0.1,0) rectangle ++(-0.25,0.5);
	\draw[line width=0.75pt] (-0.85,0) rectangle ++(-0.25,0.5); 
	\draw[line width=0.75pt] (-1.35,0.5) rectangle ++(-0.25,0.5);
		
	\draw (0.35,0.25) -| (0.45,1.25) node[above] {$B_{t-1}$};
	\draw (0.45,0.75) -- (1.35,0.75);
	\draw (0.975,0.5) -- (0.975,0.75);
	\draw (1.1,0.25) -| (1.475,0.5);
	\draw (0.975,0) -- (0.975,-0.25) node[below] {$\phi_0$};
	\draw (1.475,0.25) -| (1.475,-0.5) node[below] {$B_t$};
	\draw (1.475,1) -| (1.475,1.25) node[above] {$\phi_1$};
	
	\draw (0.225,0) |- (-0.225,-0.1) -- ++(0,0.1);
	\draw (0,-0.1) -- (0,-0.5)node[below]{$(A+B)_{t}$};
	\draw (0.225,0.5) |- (-0.225,0.6) -- ++(0,-0.1);
	\draw (0,0.6) -- (0,.85)node[above]{$\phi_1$};

	\draw (-0.35,0.25) -| (-0.45,1.25) node[above] {$A_{t-1}$};
	\draw (-0.45,0.75) -- (-1.35,0.75);
	\draw (-0.975,0.5) -- (-0.975,0.75);
	\draw (-1.1,0.25) -| (-1.475,0.5);
	\draw (-0.975,0) -- (-0.975,-0.25) node[below] {$\phi_0$};
	\draw (-1.475,0.25) -| (-1.475,-0.5) node[below] {$A_t$};
	\draw (-1.475,1) -| (-1.475,1.25) node[above] {$\phi_1$};
			
	\end{tikzpicture}
		\caption{Computing OR}
		\label{fig:or_pipelined}
\end{subfigure}
		\begin{subfigure}[b]{0.49\linewidth}
			\centering
	\begin{tikzpicture}
\draw[line width=0.75pt] (0.1,0) rectangle ++(0.25,-0.5);
\draw[line width=0.75pt] (0.85,0) rectangle ++(0.25,-0.5); 
\draw[line width=0.75pt] (1.35,-0.5) rectangle ++(0.25,-0.5); 

\draw[line width=0.75pt] (-0.1,0) rectangle ++(-0.25,-0.5);
\draw[line width=0.75pt] (-0.85,0) rectangle ++(-0.25,-0.5); 
\draw[line width=0.75pt] (-1.35,-0.5) rectangle ++(-0.25,-0.5);

\draw (0.35,-0.25) -| (0.45,-1.25) node[below] {$B_{t}$};
\draw (0.45,-0.75) -- (1.35,-0.75);
\draw (0.975,-0.5) -- (0.975,-0.75);
\draw (1.1,-0.25) -| (1.475,-0.5);
\draw (0.975,0) -- (0.975,0.25) node[above] {$\phi_1$};
\draw (1.475,-0.25) -| (1.475,0.5) node[above] {$B_{t-1}$};
\draw (1.475,-1) -| (1.475,-1.25) node[below] {$\phi_0$};

\draw (0.225,0) |- (-0.225,0.1) -- ++(0,-0.1);
\draw (0,0.1) -- (0,0.5)node[above]{$(A+B)_{t-1}$};
\draw (0.225,-0.5) |- (-0.225,-0.6) -- ++(0,0.1);
\draw (0,-0.6) -- (0,-.85)node[below]{$\phi_0$};

\draw (-0.35,-0.25) -| (-0.45,-1.25) node[below] {$A_{t}$};
\draw (-0.45,-0.75) -- (-1.35,-0.75);
\draw (-0.975,-0.5) -- (-0.975,-0.75);
\draw (-1.1,-0.25) -| (-1.475,-0.5);
\draw (-0.975,0) -- (-0.975,0.25) node[above] {$\phi_1$};
\draw (-1.475,-0.25) -| (-1.475,0.5) node[above] {$A_{t-1}$};
\draw (-1.475,-1) -| (-1.475,-1.25) node[below] {$\phi_0$};

\end{tikzpicture}	
		\caption{Decomputing OR}
	\label{fig:or_pipelined_decompute}
\end{subfigure}

	\vspace*{-3mm}
	\caption{OR gate for fully-pipelined circuits}	
	\vspace*{-3mm}
\end{figure}

As for retractile circuits, we again map the OIG nodes to an adiabatic realizations of an OR gate. As mentioned above, this requires to realize each OR gate as shown in Fig.~\ref{fig:or_pipelined}.\footnote{Signals with fanout do not have to be buffered multiple times.}
This way, the signals from stage $s_{t-1}$ (e.g.~$A_{t-1}$ and $B_{t-1}$) serve as input to compute $(A+B)_t = A_{t-1} + B_{t-1}$.
Since $(A+B)_{t}$ is driven by clock $\phi_1$, its value is inherently latched. In fact, the input signals $A_{t-1}$ and $B_{t-1}$ are reset to 0 by the according buffers (disconnecting $\phi_1$ and $(A+B)_t$), before the clock $\phi_1$ is transitioned back to 0.

Now, in contrast to retractile circuits, new hardware is required to decompute the result (after e.g.~copying it elsewhere) since the stages of the pipeline already contain the values of the next computation. 
The (conditionally) reversible function calculated by the pipeline is $F = f_{N-1}\circ f_{N-2} \circ \cdots \circ f_0$, where $f_i$ is the conditionally reversible function computed by stage $s_i$.\footnote{Note that $\circ$ denotes functional composition, i.e.~$g(x)\circ f(x) = g(f(x))$.} Since the function $f_i$ computed by each stage is conditionally reversible, the inverse of $F$ (i.e.~$F^{-1}$) exists and is determined by $F^{-1} = f_0^{-1} \circ f_1^{-1} \circ \cdots \circ f_{N-1}^{-1}$. 
The inverse $f_i^{-1}$ of the function $f_i$ computed by stage $s_i$ can be easily realized by duplicating the hardware for stage $s_i$ and connecting the power clocks $\phi_0$ and $\phi_1$ in opposite fashion (as shown for an OR gate in~Fig.~\ref{fig:or_pipelined_decompute}).
Consequently, decomputing the results requires to double the depth of the pipeline and, thus, doubles the number of required transmission gates.

\begin{myexample}
	Consider again the circuit shown in Fig.~\ref{fig:circuit_retractile}. The first stage contains a single OR gate. Additionally, three buffers are required to forward the inputs $x_2$, $x_1$, and $x_0$ to stage $s_1$ (while decomputing them in stage $s_0$). Consequently, $(1+3)\cdot 2=8$ transmission gates are required. The second stage has four input signals and requires two OR gates. Therefore, $(4+2)\cdot 2 = 12$ transmission gates are required to realize stage $s_1$. The third stage has then 6 inputs and requires $16$ transmission gates. Finally, the last stage has 8 inputs and requires $20$ transmission gates. Overall, this sums up to $56$ transmission gates. The reverse cascade of the stages again requires 56 transmission gates. Consequently, a total of 112 transmission gates are required (448 if we take the quad-rail encoding into account) to realize the function in a fully-pipelined fashion---a huge overhead compared to the retractile design methodology. However, the circuit has a higher throughput and only requires two different clocks to be operated (four if we take into account that their complement is also needed due to a dual-rail encoding).
\end{myexample}

\subsection{Advanced Solution}
\label{sec:pipelined_opt}

The mapping scheme discussed above yields circuits with a huge overhead since many signals are pushed through the whole pipeline---even though they are not required as outputs, nor to obtain reversibility of a stage. Hence, we propose to decompute such unnecessary signals as soon as possible. 
As shown in Fig.~\ref{fig:or_pipelined_decompute}, the inputs of a gate have to be present until its output is decomputed.
This means, the signals resulting from the gates in the next-to-last stage can be decomputed while computing the outputs of the function to be realized. Afterwards, the signals generated in the stage before can be decomputed---eventually resulting in the mapping scheme discussed in the previous subsection---hence, no signal can be decomputed before the final outputs of the function to be realized are determined.

However, we can easily circumvent this problem by choosing some signals that shall not be decomputed.\footnote{Note that, in the end, all signals are decomputed since each stage is duplicated as discussed in Section~\ref{sec:pipelined_sf}.} 
To this end, we mark the corresponding OIG nodes that generate these signals. This allows to decompute several other signals earlier---while continuing to compute the outputs of the function. Consequently, fewer signals are pushed through the pipeline---reducing the number of required transmission gates. 

Recall, that each node $v$ of the OIG is translated to an OR gate on a certain stage of the circuit.  
To determine when the signal resulting from $v$ can be decomputed we traverse all parents (denoted $p_j$ in the following). For each parent node $p_j$ we determine the stage in which the signal generated by $v$ can be decomputed at the earliest. Then, we take the stage with the largest index, since the constraints for all parents have to be satisfied.
If $p_j$ is a node that is marked, we can immediately decompute the signal generated by $v$ in the same stage (since the signal computed by $p_j$ is not decomputed afterwards).
If $p_j$ is not marked, we can decompute the signal generated by $v$ at the earliest one stage after the signal generated by $p_j$ can be decomputed (because the signal generated by $v$ is required to decompute the signal generated by $p_j$).

\begin{myexample}
	\label{ex:pipelined_opt}
Consider again the OIG shown in Fig.~\ref{fig:oig} (as well as the corresponding circuit shown in Fig.~\ref{fig:circuit_retractile}). Assume that we marked the nodes labeled $\vee_2$ and $\vee_3$ (the nodes labeled $\vee_6$ and $\vee_7$ are inherently marked since they are directly connected to an output).
Consequently, we want to decompute the signals generated by the OIG nodes labeled $\vee_1$, $\vee_4$, and $\vee_5$ as soon as possible.
In the second stage (i.e.~$s_1$) of the circuit, we compute the result of the nodes labeled $\vee_2$ and $\vee_3$. Since the signal generated by node $\vee_1$ is not required anymore (its single parent labeled $\vee_3$ is marked), it can be decomputed in the second stage as well. Consequently, we can save the buffers for this signal in the third and fourth stage of the circuit. Furthermore, the signals generated by nodes labeled $\vee_4$ and $\vee_5$ can be decomputed while computing the outputs of the function (in stage $s_3$). Since this is the last stage of the circuit, no buffers can be saved. However, fewer output signals result.
Considering the fact that each pipeline stage has to be duplicated, a reduction of four buffers (i.e.~8 transmission gates) can be obtained. 
\end{myexample}

This leads to the question how to determine a suitable marking scheme for the nodes, i.e.~a marking scheme that results in a circuit with a smaller number of transmission gates. 
A very simple but also effective marking scheme is to mark all nodes of the OIG with a depth that is a multiple of a constant $k\in \mathbb{N}$. For $k=2$, this means to mark all nodes with an even depth (as done in Example~\ref{ex:pipelined_opt}). The experimental evaluations summarized in Section~\ref{sec:exp} show that significant improvements can be obtained by using this marking scheme.

\section{Evaluation}
\label{sec:exp}

In this section, we summarize and discuss the results obtained by our evaluations of the proposed design methods for adiabatic circuits. To this end, we implemented the approaches discussed in Section~\ref{sec:retractile} and Section~\ref{sec:pipelined} in C++
and used the tool ABC~\cite{DBLP:conf/cav/BraytonM10} to generate the initially required AIGs/OIGs (to reduce the number of AIG nodes, we used the synthesis command \emph{dc2}). 
Afterwards, we evaluated the resulting methods using  benchmarks taken from the ISCAS~\cite{ISCAS:89} and the IWLS benchmark suite~\cite{McE:93}.
 
Table~\ref{tab:results} summarizes the obtained results. The first columns show the name of the benchmark as well as the number of primary inputs~\emph{PI} and primary outputs \emph{PO}.  
Then, we list the results obtained for retractile and fully-pipelined adiabatic circuits. For each design style, we list the number of required transmission gates (denoted~$\left|tg\right|$) and the number of required power clocks (denoted~$\left|\phi\right|$) of the straightforward solution as well as the advanced solution (columns denoted \emph{Straight-forward} and \emph{Advanced}, respectively). Having a dual-rail (for retractile circuits) or quad-rail encoding (for fully-pipelined circuits) is taken into account in the numbers listed for the required transmission gates, as well as the fact that each power clock has to be supplied in two polarities (i.e.~a power clock is dual-rail encoded for both types of circuits).
For sake of completeness, we also list the parameter $k$ used in the solution discussed in Section~\ref{sec:pipelined_opt}. 
The runtime is not listed in Table~\ref{tab:results} since all methods are capable to produce these results in negligible runtime (i.e.~a fraction of a second).

\begin{table}[t]
	\caption{Evaluation}
	\label{tab:results}
	\vspace*{-2mm}
	\centering
	\scriptsize
	\setlength{\tabcolsep}{3pt}
	\renewcommand{\arraystretch}{1.1}
	\begin{tabular}{lrr||rr|rr||rr|rrr}
		\multicolumn{3}{c||}{}  & \multicolumn{4}{c||}{Retractile (Section~\ref{sec:retractile})} & \multicolumn{5}{c}{Fully-pipelined (Section~\ref{sec:pipelined})} \\
		\multicolumn{3}{c||}{} & \multicolumn{2}{c|}{Straight-forward} & \multicolumn{2}{c||}{Advanced}  & \multicolumn{2}{c|}{Straight-forward} & \multicolumn{3}{c}{Advanced} \\
\multicolumn{3}{c||}{} & \multicolumn{2}{c|}{(Section~\ref{sec:retractile_sf})} & \multicolumn{2}{c||}{(Section~\ref{sec:retractile_opt})}  & \multicolumn{2}{c|}{(Section~\ref{sec:pipelined_sf})} & \multicolumn{3}{c}{(Section~\ref{sec:pipelined_opt})} \\
		Name & $PI$ & $PO$ & $\left|\phi\right|$ & $\left|tg\right|$ & $\left|\phi\right|$ & $\left|tg\right|$ & $\left|\phi\right|$ & $\left|tg\right|$ & $k$ & $\left|\phi\right|$ & $\left|tg\right|$ \\ \hline
		\begin{filecontents*}{results.csv}
file,inputs,outputs,max_depth,retractile_sf,retractile_opt,pipelined_sf,pipelined_opt,,
apex5,117,88,26,24,1826,1181,150544,2,112208
ex4p,128,28,28,20,2020,1260,205968,3,136080
o64,130,1,16,2,258,130,30928,2,23936
i3,132,6,12,2,252,132,23024,2,18528
i5,133,66,36,12,264,160,60096,4,50384
i8,133,81,30,24,1804,1079,147104,3,117504
apex6,135,99,26,20,1210,793,123104,2,87216
x3,135,99,26,18,1230,773,131120,4,88112
rot,135,107,50,34,1110,773,221936,2,161504
i6,138,67,12,8,766,461,36320,4,29984
frg2,143,139,24,20,1508,958,116496,3,92288
pair,173,137,44,40,2608,1693,373552,3,245584
c5315,178,123,56,38,2754,1722,559472,3,293152
i4,192,6,28,2,372,192,75200,3,59776
i7,199,67,12,8,1012,586,51392,4,42384
i2,201,1,26,6,416,210,79056,3,58864
c7552,207,108,76,62,2940,2085,868272,4,474912
c2670,233,140,44,20,1076,663,232496,3,152032
des,256,245,36,26,6784,4495,752016,5,499184
i10,257,224,66,48,3440,2292,776400,3,503184
\end{filecontents*}
\csvreader[
		late after line=\\,
		late after last line=\\,
		]{results.csv}
		{1=\Name,2=\In, 3=\Out, 4=\Dsf, 5=\Dopt, 6=\retractileSF, 7=\retractileOPT, 8=\pipelinedSF, 9=\kopt, 10=\pipelinedOPT}
		{\Name & \In & \Out & \Dsf & \optnum{\retractileSF} & \Dopt & \optnum{\retractileOPT} & 4 & \optnum{\pipelinedSF} & \kopt & 4 & \optnum{\pipelinedOPT}}
\end{tabular}\\

\raggedright{$\left|\phi\right|$: \#required clocks \hspace*{0.4cm} $\left|tg\right|$: \#transmission gates \hspace*{0.4cm}  $k$: parameter discussed in Section~\ref{sec:pipelined_opt}}
\vspace*{-5mm}
\end{table}

First, the results nicely show the impact of the respectively chosen design style. 
Retractile circuits are clearly the better choice when it comes to reducing the number of gates, while pipelined circuits are efficient with respect to the number of power clocks and, following that, also the throughput. At a first glance, it might look that the costs of having fewer power clocks in pipelined circuits is not acceptable  (in fact, magnitudes more gates are required). However, if area is not an issue, this might still acceptable since, as discussed in Section~\ref{sec:background},  gates in adiabatic circuits do not affect the energy consumption as much as they do in conventional circuits. 
Hence, each design style has its own advantages and disadvantages and, eventually, the user is presented with complementary solutions out of which the best suitable can be chosen.

Besides that, the results clearly show the improvement of the advanced schemes. On average an improvement of approx.~42\% in the number of required power clocks, as well as an average improvement of  approx.~37\% with respect to the number of required transmission gates is obtained for retractile circuits.
For the fully-pipelined circuits, we observe a reduction in the number of transmission gates of  approx.~30\% on average. 
Overall, these results clearly confirm the benefit and applicability of the proposed design automation techniques for this kind of circuits. While previously considered circuits were either handcrafted (following approaches e.g.~proposed in~\cite{younis1993practical,anantharam2004driving}) or relied on fully reversible realizations which led to an unnecessarily large overhead (as conducted in~\cite{DBLP:journals/tcad/MorrisonR14,rauchenecker2017exploiting} and discussed in~\cite{DBLP:conf/rc/Frank17}),
the proposed design flow allows for generating the desired adiabatic circuits in an automatic fashion while, at the same time, satisfying the switching rules by conditional reversibility only. The improvements obtained by the advanced schemes additionally show the further potential that can be exploited following this direction.

\section{Conclusions}
\label{sec:conclusions}

In this work, we proposed an automatic and dedicated design flow for adiabatic circuits which explicitly takes recent findings in this domain (namely that conditional reversibility is sufficient for adiabatic circuits) into account. The proposed flow first realizes the desired functionality in terms of an AIG/OIG and, afterwards, dedicatedly maps the resulting structure to an adiabatic description. For the latter step, two complementary schemes (namely retractile or fully-pipelined) are considered which allow the designer to either focus on reducing the number of gates or keeping the number of power clocks small. Furthermore, optimizations are proposed which allow for a reduction in the number of gates by approx.~37\% and 30\%, respectively, for both design styles on average.
By this, expertise from both, adiabatic circuits and design automation, is combined yielding an automatic \emph{and} dedicated design scheme for this promising technology. This eventually provides the basis for further studies including, besides others, more sophisticated optimizations, the design and use of larger building blocks, as well as 
the application of the proposed design flow in the physical implementation of adiabatic circuits.

\begin{acks}
This work has partially been supported by the European Union through the
COST Action IC1405. M. Frank was supported by the Laboratory Directed Research and Development program at Sandia National Laboratories and by the Advanced Simulation and Computing program under the U.S. Department of Energy’s National Nuclear Security Administration (NNSA). Sandia National Laboratories is a multimission laboratory managed and operated by National Technology and Engineering Solutions of Sandia, LLC., a wholly owned subsidiary of Honeywell International, Inc., for NNSA under contract DE-NA0003525. Approved for public release, SAND2018-9936 O.
\end{acks}

\bibliographystyle{ACM-Reference-Format}
{
	\bibliography{literature}
}
\end{document}